\documentstyle[prl,aps,epsfig]{revtex}

\begin{document}
\twocolumn
\draft
\preprint{HEP/123-qed}

\wideabs{
\title{
Single and Double Resonance Microwave Spectroscopy 
in Superfluid $^4$He Clusters} 

\author{
I. Reinhard$^*$, C. Callegari, A. Conjusteau, K.K. Lehmann and G. Scoles}
\address{Department of Chemistry, Princeton University, Princeton, NJ 08544}

\date{May 7, 1999}

\maketitle
\begin{abstract}
Purely rotational transitions of a molecule 
embedded in large $^4$He clusters have been detected for the first time.
The saturation behavior shows that, in contrast to previous expectations,
the microwave line profiles are dominated
by inhomogeneous broadening mechanisms.
Spectral holes and peaks produced by
microwave-microwave double resonance
have widths comparable to those of single resonance lines,
indicating that relaxation occurs among quantum states of
the inhomogeneous distribution of each rotational level, at a rate
$\approx$10 times
faster than rotationally inelastic relaxation.
\end{abstract}
} 

The spectroscopy of molecules embedded in large $^4$He clusters
has recently received considerable experimental and theoretical attention
\cite{Wha_96,toe_ARPC98}.
Nanometer scale helium clusters (nanodroplets), 
containing from several hundreds to more than 10$^4$ He atoms,
provide a unique environment for high resolution matrix spectroscopy where
the advantages of both conventional matrix spectroscopy and 
molecular beam spectroscopy are combined \cite{gs_sci98}.
Since these clusters will pick up molecules or atoms
that they encounter on their path without being appreciably deflected 
\cite{gs_jcp85},
they allow for a high degree of synthetic flexibility,
and in particular for the formation and stabilization of weakly bound and
unstable species \cite{gs_jpc93,toe_sci96,gs_prl96,gs_jpc98,miller_HCN}.
Evaporative cooling has been found to maintain
$^4$He nano\-droplets at a temperature of 0.4\,K \cite{BS_zfp90,toe_prl95},
well below the predicted superfluid transition temperature 
which ranges from 2.14\,K in bulk liquid helium to 1.5\,K for
clusters of only 10$^2$ He atoms \cite{KriWha_93}. 
As the perturbations imposed on the guest molecules by the helium host
are minimal, the shift and width of spectroscopic lines in $^4$He clusters
are considerably less than for traditional matrix environments 
\cite{Wha_96}. 
Furthermore, rotationally resolved spectra have been observed 
for a large variety of molecules 
\cite{toe_prl95,andre,miller,HCN,toe_sci98,Har_97,hui_jcp96}
which show the structure predicted by the gas phase symmetry of 
the molecules with, however, reduced rotational constants.
By showing that ro-vibrational spectra in $^3$He clusters 
collapse into a single line,
the weakly damped molecular free rotation present in liquid $^4$He
has recently been demonstrated to be
a direct consequence of the boson nature of $^4$He and is considered 
a microscopic manifestation of superfluidity \cite{toe_sci98}.

An important unresolved question posed by the IR spectra relates to the
physical process responsible for the line broadening
observed in ro-vibrational transitions, which ranges from
150\,MHz in the case of the R(0) line in
the $\nu_3$ fundamental in OCS \cite{Har_97} to
5.7\,cm$^{-1}$ for the case of the P(1) line in the $\nu_3$ asymmetric stretch
in H$_2$O \cite{hui_jcp96}.
For the carefully studied case of the $\nu_3$ fundamental of SF$_6$ 
\cite{toe_prl95},
the lines were found to have a Lorentzian shape 
of width $\approx$300\,MHz 
independent of the rotational transition, which led to the
suggestion that the linewidth reflected vibrational 
relaxation and/or dephasing \cite{Har_97}.  

Since He clusters remain fluid down to zero temperature
and because of the very large zero point motion of the $^4$He atoms, 
it appears natural to assume 
that the spectra of molecules seeded in this medium
should not display inhomogeneous effects 
other than contributions from the cluster size distribution 
via size-dependent frequency shifts 
which, however, have been shown to be small \cite{remark1}.
In solids, variations of local binding sites
lead to a distribution of vibrational frequencies, 
which results in inhomogeneous broadening
that dominates the linewidths at low temperature.
In contrast, in liquids local solvation fluctuations 
lead to dynamic dephasing. 
Treating the clusters as a classical liquid, one may expect
the timescale of the solvation fluctuation 
(due to the large zero point kinetic energy of the He atoms)
to be much faster than the dephasing times observed 
in most ro-vibrational spectra and hence the effect of fluctuating
solvation would likely be strongly motionally averaged \cite{Oxt_arpc81}, 
leading to homogeneous, Lorentzian lineshapes. 
The spectra of SF$_6$ \cite{toe_prl95} and CH$_3$CCH \cite{andre}, which
are well described with a free rotor Hamiltonian and
Lorentzian line shapes, seem to confirm the assumption
that the major source of line broadening is of homogeneous nature.

However, at temperatures as low as 0.4\,K in a superfluid medium 
solvation fluctuations are probably more appropriately described 
in terms of the interaction with the 
thermally populated fundamental modes of the cluster.
For typical cluster sizes (well below $N$=10$^5$) only surface
excitations (ripplons) have to be considered \cite{BS_zfp90}. 
The coupling of the molecular vibration to these modes
has been estimated and found to be too weak to explain
the observed linewidths \cite{Kevin}.

The observation of rotational structure in vibrational transitions 
suggests that
pure rotational spectroscopy, by excitation of microwave (MW) radiation,
should provide a useful probe of the rotational dynamics 
of the dopant molecules in the superfluid helium environment. 
It should be noted, that it was not clear a priori 
whether such spectra could be observed at all. 
Due to the short absorption path and low densities 
characteristic of molecular beams,
direct MW absorption measurements are not viable.
Whereas transitions in the UV and visible spectral range can be efficiently 
detected by laser induced fluorescence \cite{gs_prl96,gs_jpc98,toe},
beam depletion spectroscopy is employed to detect transitions in the
near and mid IR \cite{gs_jpc93,toe_prl95}.
In this method, photon absorption 
and subsequent relaxation of the molecular excitation energy 
leads to He atom evaporation from the cluster and
a decrease in the flux of He atoms in the droplet beam is observed.
While absorption of a single IR photon leads 
to the evaporation of hundreds or more He atoms from a droplet,
at least 10 microwave photons per droplet will need to be absorbed 
to provide sufficient energy to evaporate a single He atom 
($\sim$5\,cm$^{-1}$ \cite{str_jcp87}).
In order to produce a signal of sufficient size to be detected,
many He atoms per cluster must be evaporated. 
This requires the rotational relaxation to occur on a time scale
significantly shorter than 10\,$\mu$s.
Although a lower limit of the rotational relaxation time of the
order of hundreds of ps is established by the linewidth 
of typically 1\,GHz observed in the IR spectra \cite{andre}, 
the upper limit could be as high as tens of $\mu$s, 
in which case no signal would be observed.
The upper limit is imposed only by the fact that for all
observed IR spectra the rotational populations are fully thermalized
at the temperature of the He droplets by the time the clusters reach
the laser interaction region.

Here we report measurements of the microwave spectrum of HCCCN in He
nano\-clusters detected by the method of beam depletion spectroscopy.
HCCCN was used because of its large dipole moment (3.7\,Debye)
and its linear structure (rotational constant $B=$1.5\,GHz 
in the helium nano\-droplets \cite{andre,miller}),
which leads to strong and well resolved rotational transitions.


The molecular beam set-up will be described in detail elsewhere \cite{andre}.
Here we will give only a short summary
high\-lighting the aspects unique to the present study.
Clusters are formed in a supersonic free-jet helium expansion 
from a cold 5\,$\mu$m diameter nozzle
which, in the measurements presented here, is 
operated at 26\,K and a stagnation pressure of 100\,atm
yielding an average cluster size of $\approx$ 3\,$\cdot$\,10$^3$ atoms/cluster 
(estimated from Ref.\,\cite{toe_OUP95}).
After collimation by a conical skimmer, 
the clusters pass through a pickup cell 
containing typically 3$\cdot 10^{-4}$\,torr of the gas of interest 
and collect (on average) one foreign molecule each. 
Subsequently the clusters pass through a 10\,cm long P-band micro\-wave guide 
(nominal 12-18\,GHz) which is aligned parallel to the cluster beam.
The MW amplitude is modulated at 310\,Hz. 
The molecular beam enters and exits the wave\-guide 
through two 3\,mm holes in E-bends located at each end of the device. 
If multiple resonant photon absorption and subsequent relaxation of the
molecule-helium cluster system occurs, the beam depletion signal 
is recorded by a liquid helium cooled silicon bolometer 
using a lock-in technique. 

The microwave radiation is produced by a sweep generator
 (HP 8350B) with
a 0.01-26.5\,GHz plug-in  (HP 
%
\begin{figure}[b]
\centerline{\epsfxsize=8.9cm \epsfysize=5.4cm \epsffile{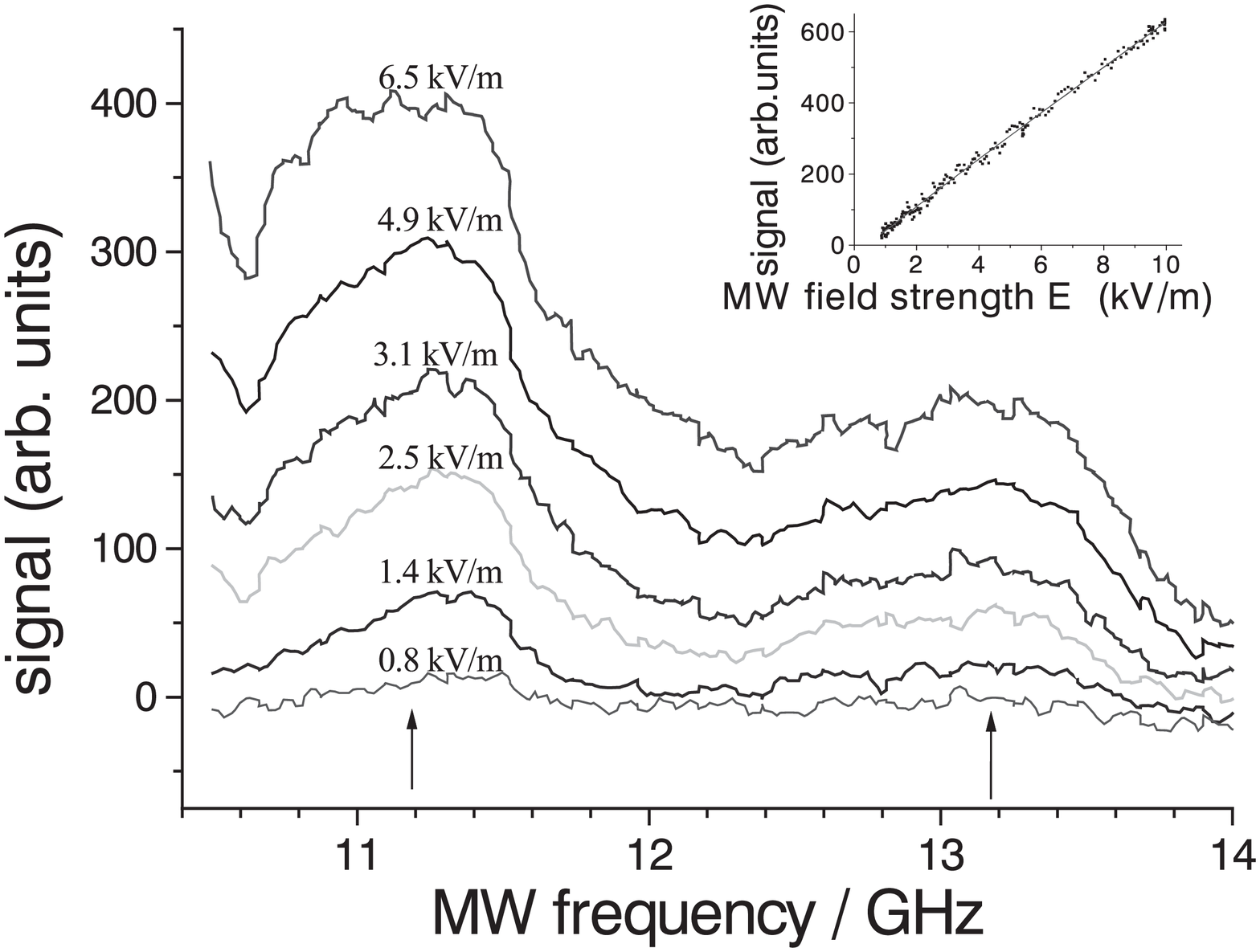}}
\unitlength 1cm
\begin{minipage}[b]{8.5cm}
    \caption{
        Microwave spectra of the J=3$\rightarrow$4 and  J=4$\rightarrow$5
        transition in HCCCN in $^4$He droplets measured at various 
        MW field strengths. The arrows indicate 
        the line positions predicted from the molecular constants 
        obtained from the ro-vibrational spectrum of HCCCN in He clusters.
        {\it Inset}: Signal amplitude of the J=4$\rightarrow$5 transition 
        at 13.1\,GHz as a function of microwave field strength. 
        The linearity of this plot demonstrates the inhomogeneous nature 
        of the dominant line broadening mechanism.}
 \end{minipage}
\end{figure}
%
\noindent 8359A) and is amplified by a
traveling wave tube amplifier (Logi Metrics A310/IJ) to a power level
between 0.05\,W and 3.4\,W (corresponding to a field strength 
of 0.78 to 6.5\,kV/m in the center of the P-band wave\-guide).
The power transmitted through the 
wave\-guide is attenuated by 30\,dB and measured by a crystal detector
(HP8473B), the output of which is used to 
level the power of the sweep generator during frequency scans.


Spectra of the J= 3$\rightarrow$4 and the J=4$\rightarrow$5 transitions 
in the ground vibrational state of HCCCN
obtained for various 
MW field strengths between 0.78 and 6.5\,kV/m are shown in Fig.\,1.
The line centers agree well with the line positions predicted 
from the molecular constants obtained from the ro-vibrational IR spectrum
of HCCCN in the helium clusters \cite{andre}.
The linewidths (FWHM) are observed to increase 
from $\sim$0.6 to $\sim$1\,GHz for the J= 3$\rightarrow$4 transition and
from $\sim$0.8 to $\sim$1.2\,GHz for the J= 4$\rightarrow$5 transition
when the micro\-wave field is increased from 0.78 to 6.5\,kV/m.
%
At low MW fields these linewidths are comparable to the width of the 
corresponding ro-vibrational transitions in the spectra 
of the fundamental CH stretching mode \cite{miller},
indicating that vibrational relaxation and dephasing
which frequently are the dominant line broadening mechanisms 
in the spectra of impurities in classical liquids \cite{Oxt_arpc81}
are not the main source of broadening 
for a molecule such as HCCCN in a superfluid helium cluster.
%
Similar linewidths have been observed by us for 
the corresponding micro\-wave transitions in CH$_3$CN and CH$_3$CCH.

The dependence of the signal amplitude $S$ on the microwave field strength $E$
has been measured for HCCCN with the MW frequency fixed at the top of either
the J=3$\rightarrow$4 or the J=4$\rightarrow$5 transition and
is well described by $S \sim  E^2/(1+E^2/E^2_{sat})^{1/2}$
with the saturation field  $E_{sat}=$1.1(2)\,kV/m (see Inset Fig.\,1).
This saturation behavior, resulting in a linear dependence of the absorption 
as function of the MW field intensity for $E \gg E_{sat}$, 
demonstrates that, in contrast to the previous expectations,
the linewidth is dominated by inhomogeneous broadening \cite{demtroeder}.
With the saturation parameter $(E/E_{sat})^2$, 
the homogeneous unsaturated linewidth is calculated to be  
at least a factor of 6 narrower than the inhomogeneous
linewidth observed at a MW field intensity of 6.5\,kV/m.
This sets the lower limit of the rotational relaxation time to about 2\,ns.

An upper limit for the rotational relaxation time has been set by
a MW amplitude modulation experiment:
With the MW frequency fixed on top of the 3$\rightarrow$4 transition
at a MW field of 7.8\,kV/m, 
the signal height is monitored while
the MW field is 100\% square wave modulated at a frequency $f$. 
By modelling the 3$\rightarrow$4 transition as a driven two level-system, 
the absorbed micro\-wave power is calculated to increase 
by a factor of 2 when $f$ changes from $f \ll 1/T_1$ to $f \gg 1/T_1$, where
$1/T_1$ is the population relaxation rate for the transition. 
This is basically independent of the dephasing rate (1/$T_2$), 
as long as the micro\-wave power is sufficiently large to allow for saturation.
From the fact that no increase in signal is observed for modulation frequencies
up to 10\,MHz we estimate an upper limit for the rotational relaxation time
of about 20\,ns. 
This limit is in agreement with the independent but less stringent
estimate inferred from the comparison of the strengths of the 
MW spectra and the IR spectra \cite{andre}, which implies that 
the rotational relaxation takes place at a rate not slower than in tens of ns. 

In order to determine the homogeneous linewidth of the rotational transition,
micro\-wave-micro\-wave double resonance experiments have been carried out. 
A second micro\-wave source (HP8690B, plug-in HP8694B: 8-12.4\,GHz) 
is employed to generate micro\-wave radiation at a fixed frequency.
%
While the first micro\-wave field (the probe) is frequency scanned 
across the 3$\rightarrow$4 and the 4$\rightarrow$5 transition,
the second micro\-wave field pumps the
J=3$\rightarrow$4 transition at about 11.1\,GHz. 

As the probe frequency approaches the pump frequency 
a strong decrease in signal is observed due to the depletion of the J=3 state
by the pump, whereas the 4$\rightarrow$5 transition signal 
is increased according to the enhanced population of the J=4 state (Fig.\,2).
Remarkably, the hole burnt into the 3$\rightarrow$4 transition
has a width of $\approx$ 50-70\,$\%$ of the single resonance linewidth,
implying that the rotational population inversion relaxation time is larger
than 4\,ns.
The increase in the 4$\rightarrow$5 signal 
even occurs over the total width of the signal,
indicating that there is a fast relaxation 
within the inhomogeneous distribution of each individual J level.
This observation is important as it 
implies that a substantial part of the inhomogeneous line broadening
is due to a dynamic effect rather than to 
\begin{figure}[t]
\centerline{\epsfxsize=8.9cm \epsfysize=5.4cm \epsffile{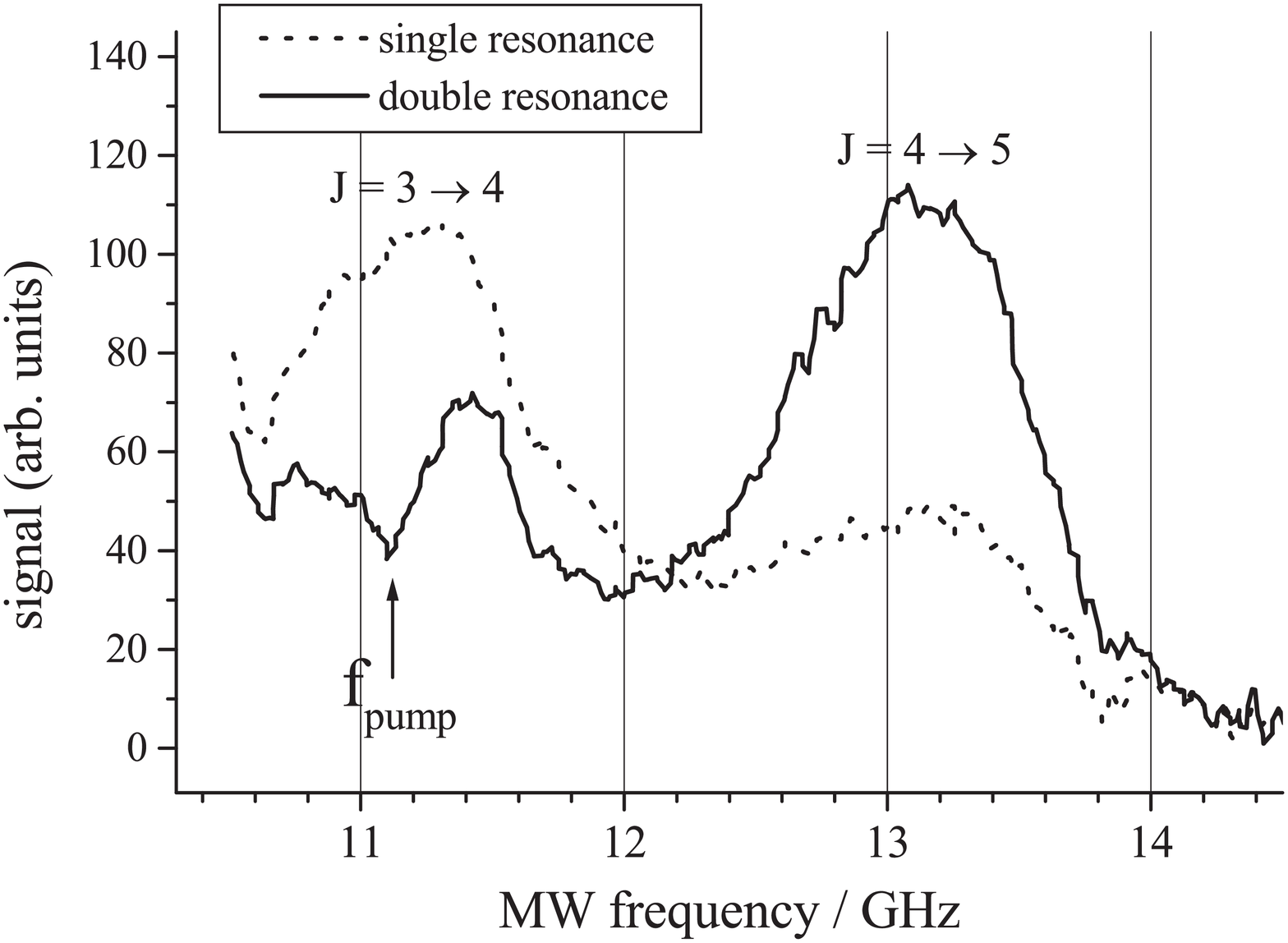}}
\unitlength 1cm
\begin{minipage}[t]{8.5cm}
%
     \caption{
        MW-MW double resonance spectrum of HCCCN in $^4$He clusters 
        compared to a single resonance spectrum.
        The depletion in the  J=3$\rightarrow$4 and 
        the enhancement in the  J=4$\rightarrow$5 transition occur over a
        significant part of the linewidths indicating rapid relaxation 
        among the substates of the inhomogeneous broadening 
        of the individual rotational levels
        with a rate much faster than the rotational population inversion rate.
        The J=3$\rightarrow$4 transition is pumped with a fixed frequency
        at 11.1\,GHz and a field of 3.75\,kV/m 
        while the probing field is 5.3\,kV/m.
}
\end{minipage}
\end{figure}
\noindent  a static effect such as the 
cluster size distribution.

The phenomenon of `dynamic' inhomogeneous broadening can be understood
assuming that there are additional degree(s) of freedom associated with
a splitting of the rotational state into several substates
and that the molecule may transit among these substates. 
Such transitions, which may well change the
kinetic and potential energy of the molecule, but produce only a small
change in its rotational energy, will be denoted as ``elastic''.
If the elastic relaxation rate
is much less than the spectral line width, 
the lineshape reflects the distribution of resonance frequencies of the
molecules in these additional quantum states. Each substate has a homogeneous
width much narrower than the width of the inhomogeneous line.
%
A double resonance experiment would be expected to show 
a correspondingly narrow hole in the pumped transition and a peak 
on top of the transition starting from the population enhanced rotor level. 
However, if the relaxation rate for the rotor quantum number
is slower still than the relaxation between the substates,
then the population disequilibrium produced by the MW pumping will be
spread over many or all the substates. This 
produces, in our case, a broad depletion in the lower rotor level
and an enhancement over the complete width of the higher rotor level.

The relative areas of the depletion and enhancement signals 
compared to the single resonance MW signal can be used to 
extract the relative rates of the two relaxation processes. 
By kinetic modelling of the transition rates between the J=3 and J=4
rotational levels and among the substates of each individual rotational state
we estimate that the elastic relaxation within one rotational state
is about one order of magnitude faster than the inelastic
population inversion relaxation.

In order to determine possible mechanisms underlying 
the observed inhomogeneous broadening one of the authors has analyzed
the dynamics of a neutral impurity in a nanometer scale $^4$He cluster 
\cite{Kevin}
showing that for an anisotropic impurity significant sources 
of line broadening arise from the coupling of the molecular rotation
with the center of mass motion of the dopant.
These couplings arise both from an anisotropic effective potential for the
dopant when shifted from the exact center of the clusters, and
from an orientationally dependent hydrodynamic contribution to the effective
inertial mass of the dopant.
It should be noted, however, that it is not clear
how the molecular energy is transferred to the cluster
since the energy released or absorbed by the molecular
rotationally inelastic or elastic transition in general is not likely to match
the quantized energy of the lowest cluster excitations.

The measurements presented here are
the first observation of a purely rotational spectrum of any
molecule in a liquid He environment, and have provided a unique window
on the sources of line broadening and in particular
onto the rotational dynamics of the dopant 
in the superfluid helium environment. 
It has been unambiguously demonstrated that
the rotational lines are dominated by inhomogeneous broadening
which is attributed to the coupling of 
the center of mass motion of the molecule
within the finite size cluster to the molecular rotation.
The second major observation is that the molecule transits among 
the quantum states of the inhomogeneous distribution
on a time scale much faster than the rotational relaxation. 
These MW-MW double resonance measurements have been followed by
a separate study using micro\-wave - infrared double resonance 
which provides new information
on the relaxation dynamics of the dopant in the cluster
and its dependence on the finite cluster size \cite{MW_IR}.
%

We would like to acknowledge R.E. Miller and his coworkers for
the free flow of information between the two groups.
We are indebted to Prof. W. Warren and Prof. S. Staggs
for providing us with the MW sweep generators
and to Dr. J. Fraser for lending us the traveling wave tube amplifier. 
This work was supported by the National Science Foundation (CHE-97-03604). 
I.R. is grateful to the Alexander-von-Hum\-boldt Foundation for financial support.

\bigskip


%



{\footnotesize
$^*$ Present address: 
          Physikalisches Institut, Universit\"at Heidelberg, Germany}
\bibliographystyle{prsty}

\begin{thebibliography}{10}

\bibitem{Wha_96}        K.B.~Whaley,
                        ``Spectroscopy and Microscopic Theory 
                        of Doped Helium Clusters'',
                        in: Advances in Molecular Vibrations and 
                        Collision Dynamics,
                        Vol.III, Ed. J.~Bowman, JAI Press Inc. (1997)

\bibitem{toe_ARPC98}    J.P. Toennies, A. Vilesov, 
                        Ann. Rev. Phys. Chem. {\bf49}, 1 (1998)

\bibitem{gs_sci98}      K.K.~Lehmann and G.~Scoles, 
                        Science {\bf279}, 2065 (1998)

\bibitem{gs_jcp85}      T.E.~Gough, M.~Mengel, P.A.~Rowntree, G.~Scoles,
                        J. Chem. Phys. {\bf83}, 4958 (1985)




\bibitem{gs_jpc93}      S.~Goyal, D.L.~Schutt, G.~Scoles,
                        J. Chem. Phys. {\bf97}, 2236 (1993)

\bibitem{toe_sci96}     M.~Hartmann, R.~Miller, J.P.~Toennies, A.F.~Vilesov,
                        Science {\bf272}, 1631 (1996)

\bibitem{gs_prl96}      J.~Higgins, W.E.~Ernst, C.~Calle\-gari, J.~Reho,
                        K.K.~Leh\-mann, G.~Scoles, M.~Gutowski,
                        Phys. Rev. Lett. {\bf77}, 4532 (1996);
                        J.~Higgins, C.~Calle\-gari, J.~Reho, F.~Stienke\-meier,
                        W.E.~Ernst, K.K.~Lehmann, M.~Gutowski, G.~Scoles,
                         Science {\bf273}, 629 (1996)

\bibitem{gs_jpc98}      
                        J.~Higgins, C.~Calle\-gari, J.~Reho, F.~Stienke\-meier,
                        W.E.~Ernst, M.~Gutowski, G.~Scoles,
                        J. Phys. Chem. A {\bf102}(26), 4952 (1998)


\bibitem{miller_HCN}    K. Nauta, R.E. Miller, Science {\bf283}, 1895 (1999)
\bibitem{BS_zfp90} D.M. Brink, S. Stringari, Z. Phys. D {\bf15}, 257 (1990)

\bibitem{toe_prl95}     M. Hartmann, R.E.~Miller, J.P.~Toennies, A.~Vilesov,
                        Phys. Rev. Lett. {\bf75}, 1566 (1995)


\bibitem{KriWha_93} 
                        M.V. Rama Krishna, K.B. Whaley,
                        ``Superfluidity in Helium Clusters'', p.257,
                        in: On Clusters and Clustering, Ed. P.J. Reynolds,
                        North-Holland (1993)


\bibitem{andre}         A. Conjusteau, C. Callegari, I. Reinhard,
                        K.K. Lehmann, G. Scoles, to be published 

\bibitem{miller}        K. Nauta, R.E. Miller, private communication  

\bibitem{HCN}           C. Callegari, A. Conjusteau, I. Reinhard,
                        K.K. Lehmann, G. Scoles,
                        K. Nauta, R.E. Miller, to be published 

\bibitem{toe_sci98}     S.~Grebenev, J.P.~Toennies, A.F.~Vilesov,
                        Science {\bf279}, 2083 (1998)


\bibitem{Har_97}        M.~Hartmann, doctoral thesis, G\"ottingen (1997)

\bibitem{hui_jcp96}     R.~Fr\"ochtenicht, M.~Kaloudis, M.~Koch, F.~Huisken,
                        J. Chem. Phys. {\bf105}, 6128 (1996)

\bibitem{remark1}
        From the analysis of the IR spectra 
        as a function of mean cluster size,
        it can be concluded that for large clusters (N$\approx$3000), 
        the cluster size distribution 
        should produce a broadening of $<$200\,MHz in the IR spectrum,
        which is just a small fraction of the observed linewidth
        of $\approx$1\,GHz, though it can be expected to `wash out' any high
        resolution features in the spectrum.

\bibitem{Oxt_arpc81}    
                        D.W. Oxtoby, Ann. Rev. Phys. Chem. 1981, 77-101

\bibitem{Kevin}         K.K.~Lehmann, accepted for publ. at
                        Mol. Phys. (1998)


\bibitem{toe}           M. Hartmann, F. Mielke, J.P. Toennies, 
                        A.F Vilesov, G. Benedek,
                        Phys. Rev. Lett {\bf76}, 204 (1997); 
                        M. Hartmann, A. Lindinger, J.P. Toennies, A.F. Vilesov,
                        Chem. Phys. {\bf239}, 139 (1998)

\bibitem{str_jcp87}     S. Stringari, J. Treiner, 
                        J. Chem. Phys. {\bf87}, 5021 (1987)
        

\bibitem{toe_OUP95}     E.L. Knuth, B. Schilling, J.P. Toennies,
                        ``On Scaling Parameters for Predicting Cluster Sizes 
                        in Free Jets'',
                        in: Internat. Symp. on Rarefied Gas Dynamics, p.270,
                        Oxford University Press (1995)

\bibitem{demtroeder}    see e.g.: W. Demtr\"oder, Laser Spectroscopy, 2nd ed.,
                        (Springer, Berlin, Heidelberg, New York 1996),
                        chap. 7.2

\bibitem{MW_IR}         C. Callegari, I. Reinhard, K.K. Lehmann, G. Scoles,
                        K. Nauta, R.E. Miller, to be published

\end{thebibliography}
    
\end{document}